\documentclass{aa} 

\usepackage{color}
\usepackage{graphicx}
\usepackage{natbib,twoopt}
\usepackage[varg]{txfonts}
\usepackage{gensymb}
\usepackage{lscape}

\bibpunct{(}{)}{;}{a}{}{,}

\makeatletter\renewcommand\theenumi{\@alph\c@enumi}\makeatother

\def\hi{\relax \ifmmode {\mbox H\,\texjtsc{i}}\else H\,{\scshape i}\fi}
\def\hii{\relax \ifmmode {\mbox H\,\textsc{ii}}\else H\,{\scshape ii}\fi}
\def\nii{\relax \ifmmode {\mbox N\,\textsc{ii}}\else N\,{\scshape ii}\fi}
\def\oi{\relax \ifmmode {\mbox O\,\textsc{i}}\else O\,{\scshape i}\fi}
\def\oii{\relax \ifmmode {\mbox O\,\textsc{ii}}\else O\,{\scshape ii}\fi}
\def\oiii{\relax \ifmmode {\mbox O\,\textsc{iii}}\else O\,{\scshape iii}\fi}
\def\cii{\relax \ifmmode {\mbox C\,\textsc{ii}}\else C\,{\scshape ii}\fi}
\def\ciii{\relax \ifmmode {\mbox C\,\textsc{iii}}\else C\,{\scshape iii}\fi}
\def\civ{\relax \ifmmode {\mbox C\,\textsc{iv}}\else C\,{\scshape iv}\fi}
\def\hei{\relax \ifmmode {\mbox He\,\textsc{i}}\else He\,{\scshape i}\fi}
\def\heii{\relax \ifmmode {\mbox He\,\textsc{ii}}\else He\,{\scshape ii}\fi}
\def\mgii{\relax \ifmmode {\mbox Mg\,\textsc{ii}}\else Mg\,{\scshape ii}\fi}
\def\sii{\relax \ifmmode {\mbox S\,\textsc{ii}}\else S\,{\scshape ii}\fi}
\def\neiii{\relax \ifmmode {\mbox Ne\,\textsc{iii}}\else Ne\,{\scshape iii}\fi}
\def\ariv{\relax \ifmmode {\mbox Ar\,\textsc{iv}}\else Ar\,{\scshape iv}\fi}
\def\ni{\relax \ifmmode {\mbox N\,\textsc{i}}\else N\,{\scshape i}\fi}
\def\ariii{\relax \ifmmode {\mbox Ar\,\textsc{iii}}\else Ar\,{\scshape iii}\fi}
\def\caii{\relax \ifmmode {\mbox Ca\,\textsc{ii}}\else Ca\,{\scshape ii}\fi}

\usepackage[colorlinks]{hyperref} 
\hypersetup{colorlinks = true, citecolor = blue}

\begin{document}


   \title{Arm and interarm abundance gradients in CALIFA spiral galaxies}
   \titlerunning{Arm and interarm abundance gradients in CALIFA spiral galaxies}
   
   \author{L.~S\'anchez-Menguiano\inst{1,2}\and S.~F.~S\'anchez\inst{3}\and I.~P\'erez\inst{2,4}\and V.~P.~Debattista\inst{5}\and T.~Ruiz-Lara\inst{2,4,6,7}\and E.~Florido\inst{2,4}\and O.~Cavichia\inst{8}\and L.~Galbany\inst{9}\and R.~A.~Marino\inst{10}\and D.~Mast\inst{11,12}\and P.~S\'anchez-Bl\'azquez\inst{13}\and J.~M\'endez-Abreu\inst{6,7,14}\and A.~de~Lorenzo-C\'aceres\inst{3}\and C.~Catal\'an-Torrecilla\inst{15}\and M.~Cano-D\'iaz\inst{3}\and I.~M\'arquez\inst{1}\and D.~H.~McIntosh\inst{16}\and Y.~Ascasibar\inst{17,18}\and R.~Garc\'ia-Benito\inst{1}\and R.~M.~G\'onzalez~Delgado\inst{1}\and C.~Kehrig\inst{1}\and \'A.~R.~L\'opez-S\'anchez\inst{19,20}\and M.~Moll\'a\inst{21}\and J.~Bland-Hawthorn\inst{22}\and C.~J.~Walcher\inst{23}\and L.~Costantin\inst{24}}
   \authorrunning{L.~S\'anchez-Menguiano et al.}

   \institute{Instituto de Astrof\'isica de Andaluc\'ia (CSIC), Glorieta de la Astronom\'ia s/n, Aptdo. 3004, E-18080 Granada, Spain\\
              \email{lsanchez@iaa.es}
         \and Dpto. de F\'isica Te\'orica y del Cosmos, Universidad de Granada, Facultad de Ciencias (Edificio Mecenas), E-18071 Granada, Spain
         \and Instituto de Astronom\'ia, Universidad Nacional Aut\'onoma de M\'exico, A.P. 70-264, 04510, M\'exico, D.F.
         \and Instituto Carlos I de F\'isica Te\'orica y computacional, Universidad de Granada, E-18071 Granada, Spain
         \and Jeremiah Horrocks Institute, University of Central Lancashire, Preston, PR1 2HE, UK
         \and Instituto de Astrof\'isica de Canarias, Calle V\'ia L\'actea s/n, E-38205 La Laguna, Tenerife, Spain
         \and Universidad de La Laguna, Dpto. Astrof\'isica, E-38206 La Laguna, Tenerife, Spain
         \and Instituto de F\'isica e Qu\'imica, Universidade Federal de Itajub\'a, Av. BPS, 1303, 37500-903, Itajub\'a-MG, Brazil
         \and PITT PACC, Department of Physics and Astronomy, University of Pittsburgh, Pittsburgh, PA 15260, USA 
         \and Institute for Astronomy, Department of Physics, ETH Z\"urich, Switzerland
         \and Observatorio Astron\'omico, Laprida 854, X5000BGR, C\'ordoba, Argentina
         \and CONICET, Avda. Rivadavia 1917, C1033AAJ, CABA, Argentina
         \and Instituto de Astrof\'isica, Pontificia Universidad Cat\'olica de Chile, Av. Vicu\~na Mackenna 4860, 782-0436 Macul, Santiago, Chile
         \and School of Physics and Astronomy, University of St. Andrews, SUPA,
North Haugh, KY16 9SS, St. Andrews, UK
         \and Departamento de Astrof\'isica y CC. de la Atm\'osfera, Universidad
Complutense de Madrid, 28040 Madrid, Spain
         \and  Department of Physics \& Astronomy, University of Missouri-Kansas City, Kansas City, MO 64110, USA
         \and Departamento de F\'{i}sica Te\'{o}rica, Universidad Aut\'{o}noma de Madrid, Madrid 28049, Spain
         \and Astro-UAM, UAM, Unidad Asociada CSIC
         \and Australian Astronomical Observatory (AAO), PO Box 915, North Ryde, NSW 1670, Australia
         \and Department of Physics and Astronomy, Macquarie University, NSW 2109, Australia
         \and CIEMAT, Avda. Complutense 40, E-28040 Madrid, Spain
         \and Sydney Institute for Astronomy, School of Physics A28, University of Sydney,
NSW 2006, Australia.
         \and Leibniz-Institut f\"ur Astrophysik Potsdam (AIP), An der Sternwarte 16, D-14482 Potsdam, Germany. 
         \and Dipartimento di Fisica e Astronomia `G. Galilei', Universit\`a di Padova, vicolo dell'Osservatorio 3, I-35122 Padova, Italy\\
             } 

   \date{Received 15 November 2016 / Accepted 6 May 2017}


\abstract{Spiral arms are the most singular features in disc galaxies. These structures can exhibit different patterns, namely grand design and flocculent arms, with easily distinguishable characteristics. However, their origin and the mechanisms shaping them are unclear. The overall role of spirals in the chemical evolution of disc galaxies is another unsolved question. In particular, it has not been fully explored if the \hii\,regions of spiral arms present different properties from those located in the interarm regions. Here we analyse the radial oxygen abundance gradient of the arm and interarm star forming regions of 63 face-on spiral galaxies using CALIFA Integral Field Spectroscopy data. We focus the analysis on three characteristic parameters of the profile: slope, zero-point, and scatter. The sample is morphologically separated into flocculent versus grand design spirals and barred versus unbarred galaxies. We find subtle but statistically significant differences between the arm and interarm distributions for flocculent galaxies, suggesting that the mechanisms generating the spiral structure in these galaxies may be different to those producing grand design systems, for which no significant differences are found. We also find small differences in barred galaxies, not observed in unbarred systems, hinting that bars may affect the chemical distribution of these galaxies but not strongly enough as to be reflected in the overall abundance distribution. In light of these results, we propose bars and flocculent structure as two distinct mechanisms inducing differences in the abundance distribution between arm and interarm star forming regions.}


\maketitle

\section{Introduction}
\label{sec:intro}

The spiral structure is one of the most characteristic features of disc galaxies. In some cases, these patterns are well-defined, symmetric and continuous and are referred to as ``grand design'' spirals. In other cases, the spiral structure presents less symmetry and is formed of patchy arms that fade over the gaseous disc, leading to ``flocculent'' galaxies \citep{elmegreen1981}. This flocculent versus grand design nature of spiral galaxies must be directly connected to the mechanisms responsible for generating the spiral structure. However, there is no widely accepted theory that can explain the origin of this diversity of spiral arms. Although there is a general agreement that spirals are caused by density variations in the discs driven by gravitational instabilities, theories diverge in aspects such as the lifetime of these features or the mechanisms generating them. Some theories propose long-lived quasi-stationary patterns \citep[e.g.][]{linshu1964,bertinlin1996}, while others support the idea that spiral arms are short-lived, recurrent transient patterns that form through local instabilities which are swing amplified into spiral arms \citep[e.g.][]{juliantoomre1966,toomre1990}. In the current interpretation, flocculent arms are typically associated with local instabilities, whereas grand design galaxies are linked to the steady state density wave theory \citep[e.g.][]{dobbs2008,dobbs2014}. However, this question is far from being resolved.

Furthermore, this spiral classification based on flocculent or grand design systems is not always unambiguous. Multi-band observations have shown that galaxies can exhibit characteristics of both flocculent and grand design structure. In particular, some galaxies present grand design arms in the infrared (dust and old stars) and a more flocculent structure in the optical (gas and young stars) bands \citep[e.g.][]{block1991, thornley1996, thornley1997, elmegreen1999, kendall2011}. This finding makes it even more complicated to develop theories that can explain the nature of spiral structure.

\begin{figure*}
\begin{center}
\resizebox{0.75\hsize}{!}{\includegraphics{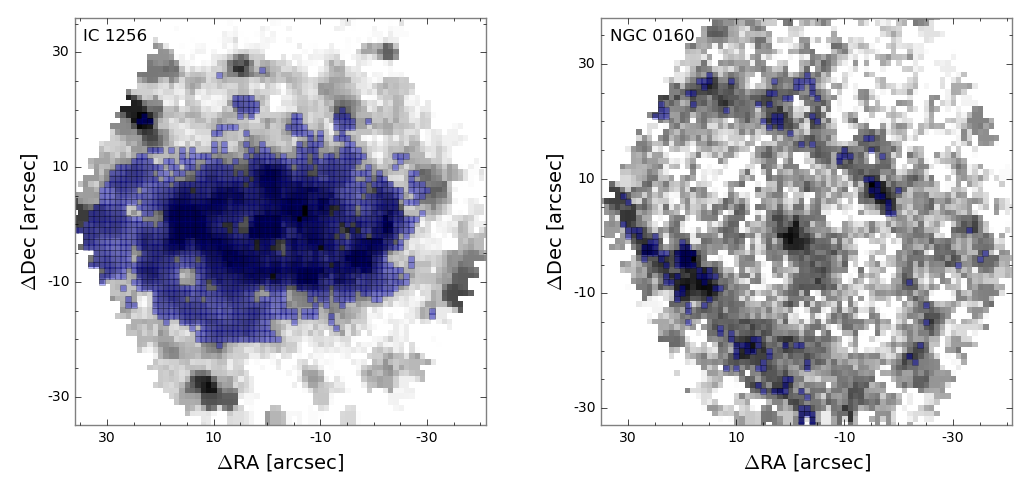}}
\caption{Location of the spaxels associated with SF regions (blue squares) superimposed on the IFS-based H$\alpha$ map derived for two galaxies of the SM16 sample, one included in this work's sample (IC~1256, left), and another that was discarded because of poor spatial coverage of the SF spaxels across the galactic disc (NGC~0160, right). See Sect.~\ref{sec:analysis1b} for an explanation of the procedure used to select the spaxels associated with SF regions.}
\label{fig:Hamaps}
\end{center}
\end{figure*}

Regardless of their origin, it is clear that spiral arms are structures where the star formation is enhanced, which may affect the chemical composition of these structures and can produce differences between the arm and interarm regions. However, only a few works have studied arm-interarm abundance variations. \citet{martin1996} analysed a sample of seven galaxies, but did not observe significant variations in the gas metallicity between these two distinct regions. \citet{cedres2002} carried out a study for two galaxies analysing different properties of the \hii\,regions (distribution, H$\alpha$ equivalent width, metallicity) of the arm and interarm regions without observing significant differences in their properties. \citet{cedres2012} studied the two-dimensional (2D) distribution of the gas metallicity of another two galaxies and found that the presence of azimuthal inhomogeneities depends on the calibrator used to derive the metallicity. On a larger scale, a few other studies have analysed possible azimuthal variations in the chemical abundance distribution of spiral galaxies \citep{kennicutt1996,rosalesortega2011,li2013}, but did not find variations related to the presence of the arms. Recently, \citet{lopezsanchez2015} found radial and azimuthal variations in the metallicity of star-forming (SF) regions along the spiral arms of NGC~1512, but these were induced by the interaction of the blue compact dwarf galaxy NGC~1510 in one of the arms.  Along these lines, \citet{zinchenko2016} analysed 88 late-type galaxies from the CALIFA second data release (DR2, \citealt{garciabenito2015}) to study the global azimuthal abundance asymmetry of these galaxies, without finding significant differences in the abundance distribution. More recently, \citet{sanchezmenguiano2016b} studied the gas content of NGC~6754 using MUSE data and found residual H$\alpha$ line-of-sight velocity and gas abundance inhomogeneities compatible with what is expected due to radial migration driven by the spiral arms \citep{grand2016}. A recent work by \citet{croaxall2016} also presents evidence of azimuthal variations which can be inferred from an intrinsic dispersion in the O/H and N/O abundance gradients.

One of the main caveats of most of these studies (with the exception of \citealt{zinchenko2016}) to provide statistically significant conclusions is the low number of analysed objects. In this paper we present an extended study of a large sample of galaxies from the CALIFA survey \citep{sanchez2012a} that allows us to derive statistically significant results, following a different approach from \citet{zinchenko2016}. For these galaxies, we analyse the oxygen abundance gradient of SF regions for both the spiral arm and the interarm areas. The main goal of this analysis is to determine if there are differences in the abundance distribution of the ionised gas associated with both structures. In addition, we also distinguish between flocculent and grand design galaxies in order to shed some light onto the origin of the spiral structure. To increase the number statistics and properly map the spatially resolved chemical abundance of the gas phase, we have carried out this study analysing the full 2D information provided by the CALIFA data avoiding any binning schemes \citep[following][hereafter SM16]{sanchezmenguiano2016}. At the physical resolution of the CALIFA data \citep{mast2014}, SM16 has already shown that a spaxel-by-spaxel analysis leads to equivalent results to those obtained following the classical procedure of detecting \hii\,regions (see also Sect.~\ref{sec:analysis1c}), with the advantages previously mentioned (i.e. an improvement in the statistics and in the spatial coverage of the gas properties). 

The paper is organised as follows. Section~\ref{sec:sample} provides a description of the sample and data we use in this study. Section~\ref{sec:analysis} describes the analysis required to derive the oxygen abundance gradients for the arm and interarm regions, including the procedure followed to separate the information of both structures. Finally, the presentation of the results, the discussion and the main conclusions are given in Sects.~\ref{sec:results} and \ref{sec:discussion}.


\section{Data and galaxy sample}\label{sec:sample}

The analysed data were collected by the CALIFA survey \citep{sanchez2012a} using the PMAS/PPAK spectrograph \citep{roth2005,kelz2006} at the 3.5m telescope of the Calar Alto observatory. The observations were carried out following a three-pointing dithering scheme designed to increase the spatial resolution, obtaining a covering factor of $100 \%$ along the full field of view of $74''\rm{x}\, 64''$. Two different set-ups were chosen for the observations: V500, with a nominal resolution of $\lambda/\Delta\lambda \sim 850$ at 5000 \AA\, (FWHM $\sim 6\,$\AA) and a wavelength range from 3745 to 7500 \AA, and V1200, with a better spectral resolution of $\lambda/\Delta\lambda \sim 1650$ at 4500 \AA\, (FWHM $\sim 2.7\,$\AA) and ranging from 3650 to 4840 \AA. The data analysed here have been reduced with version 1.5 of the reduction pipeline \citep[corresponding to the second data release,][]{garciabenito2015} providing datacubes with a final spatial resolution of FWHM $\sim 2.5''$ ($\sim1$ kpc at the typical redshift of the galaxies) and a pixel size of $1''$.  

More detailed information about the CALIFA sample can be found in \citet{walcher2014}. The observational strategy is described in \citet{sanchez2012a}. Finally, details about the data reduction and the improvements achieved along the different versions of the pipeline are available in \citet{husemann2013}, \citet{garciabenito2015}, and \citet{sanchez2016c}.

\vspace{0.5cm}

The study carried out in this paper is based on a subset of galaxies selected from the sample defined in SM16 as part of the above mentioned project. This sample comprised 122 face-on ($i < 60 \degr$) isolated spirals, including both barred and unbarred galaxies. For details about the selection criteria, see SM16.

For a proper comparison between the abundance distribution of the SF regions belonging to the spiral arms and the interarm regions, it is essential that the analysed spaxels cover a substantial extent of the disc associated with both structures. For this purpose, we discarded from the SM16 sample those galaxies for which we do not detect ionised emission associated with star formation covering a continuous and wide area of the disc (i.e. those galaxies where the fraction of non-isolated SF spaxels is lower than 15\%). Afterwards, we apply another criterion to guarantee that a statistical characterisation of the arm and interarm regions is feasible (i.e. at least 20\% of the SF spaxels belonging to each of both structures). By applying these requirements, most of the discarded galaxies are early types (Sa-Sab), maybe due to the need of a good coverage of the ionised gas throughout the disc. However, as no dependence of the abundance gradients with the morphological type has been found \citep[e.g.][]{sanchez2014,sanchezmenguiano2016}, this should not affect the results. 

Figure~\ref{fig:Hamaps} shows an example of a galaxy included in our final sample (left panel) and a galaxy that has been discarded due to the low number of spaxels covering the interarm region (right panel). In addition, galaxies whose spiral arms could not be properly traced because they were not clearly defined were also rejected (see Sect.~\ref{sec:analysis2b} for details). These cases correspond to the most flocculent galaxies of the sample, whose fragmented and blurred arms are not easy to track. By discarding these galaxies, we may be diminishing the possible differences between flocculent and grand design galaxies. However, the qualitative results should not change.

In the end, the sample analysed in this work comprises 68 galaxies that provide a good 2D coverage of the SF regions over the disc and a good characterisation of both the arm and interarm regions. For this sample of galaxies we carry out a detailed 2D study of the gas metallicity distribution to analyse possible differences between the SF regions found in the spiral arms and those located in the interarm regions.

\begin{figure*}
\resizebox{\hsize}{!}{\includegraphics{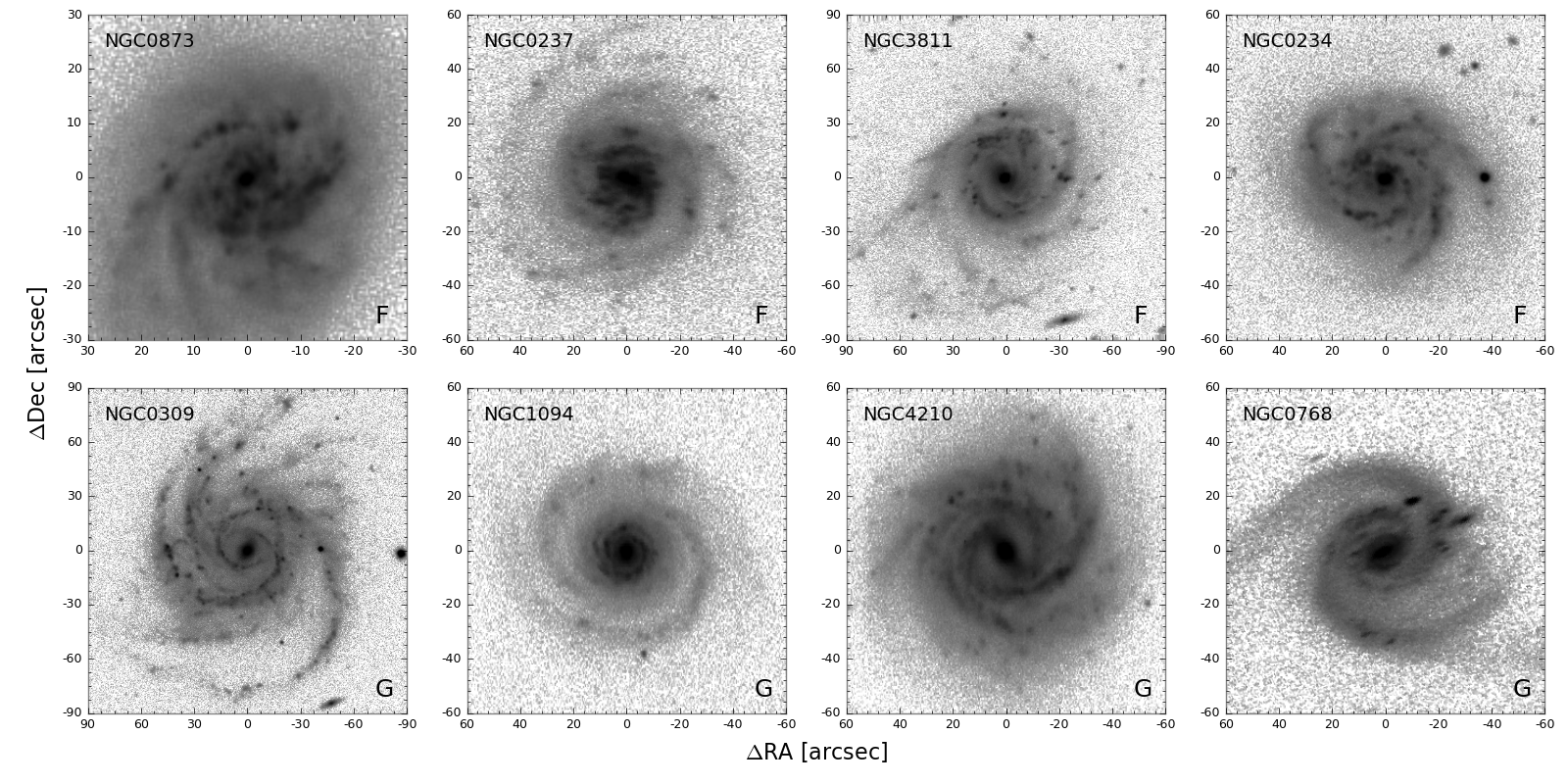}}
\caption{Deprojected SDSS $g+u$ maps for eight galaxies showing the arm classification ($F$ for flocculent and $G$ for grand design, see bottom right corner of each image).}
\label{fig:ac_mosaic}
\end{figure*}


\section{Analysis}\label{sec:analysis}

In this section we briefly summarise the procedure followed to select the spaxels, analyse their individual spectra, and derive the corresponding oxygen abundances. We refer the reader to SM16 for a more detailed explanation of this procedure. We also explain the method used to classify these spaxels according to the structure where they belong, the spiral arms or the interarm region. The subsequent derivation of the oxygen abundance gradient for both sets of spaxels is also described.

\subsection{Derivation of the oxygen abundance distribution}\label{sec:analysis1}

\subsubsection{Measurement of the emission lines}\label{sec:analysis1a}

To properly measure the flux intensity of the emission lines, the underlying stellar spectrum must be subtracted. We model both the continuum emission and the emission lines using a fitting package named FIT3D\footnote{\url{http://www.astroscu.unam.mx/~sfsanchez/FIT3D}} \citep{sanchez2011}. FIT3D fits each spectrum by a linear combination of single stellar population (SSP) templates with different ages and metallicities after correcting for the appropriate systemic velocity and velocity dispersion and taking into account the effects of dust attenuation \citep{cardelli1989}. Once the stellar component is subtracted, FIT3D proceeds to measure the emission lines performing a multi-component fitting using a single Gaussian function per emission line plus a low order polynomial function.

The entire procedure of fitting and subtracting the underlying stellar population and measuring the emission lines using FIT3D is described in more detail in \citet{sanchez2011} and \citet{sanchez2016a, sanchez2016b}.

\subsubsection{Selection of spaxels associated with star formation}\label{sec:analysis1b}

FIT3D provides us with a set of 2D intensity maps for the emission lines required to determine the gas metallicity. To guarantee realistic measurements of the emission line fluxes employed in the determination of the oxygen abundance, we discard the spaxels with emission line fluxes below $1\,\sigma$ over the continuum level (except for H$\alpha$, for which a fixed limit of $\rm 0.2 \cdot 10^{-16} \;erg\,s^{-1}\,cm^{-2}$ is imposed in order to ensure that all the selected spaxels are associated with ionised gas). Although the selected lower limit is $1-\sigma$ per spaxel, this additional requirement on the H$\alpha$ flux implies that most of the spaxels present a signal-to-noise ratio (S/N) higher than 3 for all the lines involved in the analysis (62\% for [\oiii]~$\lambda5007$, 91\% for [\nii]~$\lambda6584$, 88\% for H$\beta$, and 100\% for H$\alpha$). In principle, the reliability of the derived oxygen abundance for each spaxel, and thus that of the derived gradients, might improve with a higher S/N cut. However, a general 3-$\sigma$ cut in all the analysed lines hampers the recovery of the abundance gradient in some galaxies due to a poorer spatial coverage of the oxygen distribution, especially for the outermost interarm regions. For this reason we prefer to use the 1-$\sigma$ flux cut as this improves the statistical significance of the results. So far this is the best compromise between having a good S/N for all the emission lines and a good coverage of the oxygen abundance distribution together with a large number of analysed galaxies. We discuss the implications of the selected S/N criterion in greater detail in Sect~\ref{sec:discussion}.

This way, from all the spaxels with flux values above this 1-$\sigma$ limit, we select those that are associated with SF using well-known diagnostic diagrams. In particular, we use the diagram proposed by \citet[][]{baldwin1981} that makes use of the \mbox{[\nii]~$\lambda6584$/H$\alpha$} and \mbox{[\oiii]~$\lambda5007$/H$\beta$} line ratios, and the \citet{kewley2001} demarcation line that separates regions ionised via SF or by AGNs. 

Nevertheless, this procedure presents some difficulties in distinguishing between low-ionisation sources (weak AGNs, shocks and/or post-AGBs stars). To deal with these weak sources, we adopt the WHAN diagram \citep[W$_{\rm H\alpha}$ versus \mbox{[\nii]/H$\alpha$},][]{cidfernandes2011}, based on the use of the EW(H$\alpha$). However, we are more restrictive in the EW range, using a limit of 6 \AA\, to guarantee a higher S/N for the emission lines of all spaxels and significantly reduce the fractional contribution coming from the diffuse nebular emission \citep[ionized by the old stellar population, e.g.~][]{kehrig2012,papaderos2013,gomes2015c,morisset2016}.

\subsubsection{Derivation of the oxygen abundance distribution}\label{sec:analysis1c}

The spatial resolution of CALIFA data, at the average redshift of the survey, is of the order of a hundred of a few hundred parsecs, the size of a typical \hii\,region \citep[][]{gonzalezdelgado1997, oey2003, lopez2011}, and therefore we do not resolve the regions, allowing us to make use of strong-line empirical calibrators to measure the oxygen abundance distribution \citep{sanchezmenguiano2016}. We adopt the one based on the O3N2 index, first introduced by \citet{alloin1979} and modified by \citet{pettini2004}: 

\begin{equation}
{\rm O3N2} =  \log\left(\frac{[\oiii] \lambda5007}{{\rm H}\beta} \times \frac{{\rm H}\alpha}{[\nii] \lambda6584}\right)
\end{equation}

For this index, we use the calibration proposed by \citet{marino2013}, where \mbox{$12+\log\left({\rm O/H}\right) = 8.533 - 0.214 \,\times\, {\rm O3N2}$}. This calibration uses $T_e$-based abundances of $\sim 600$ \hii\, regions from the literature together with new measurements from the CALIFA survey, providing the most accurate calibration to date for this index. The derived abundances have a calibration error of $\pm 0.08$ dex, and the typical value associated with the pure error propagation in the measured emission lines is about 0.05 dex.

We note that $\rm T_e$-based empirical calibrations provide results that are at least $0.2-0.4$~dex lower than strong-line methods based on photoionisation models (see \citealt{lopezsanchez2010,lopezsanchez2012} for an extended discussion).

\subsection{Characterisation of spiral arms}\label{sec:analysis2}

\begin{figure*}
\resizebox{\hsize}{!}{\includegraphics{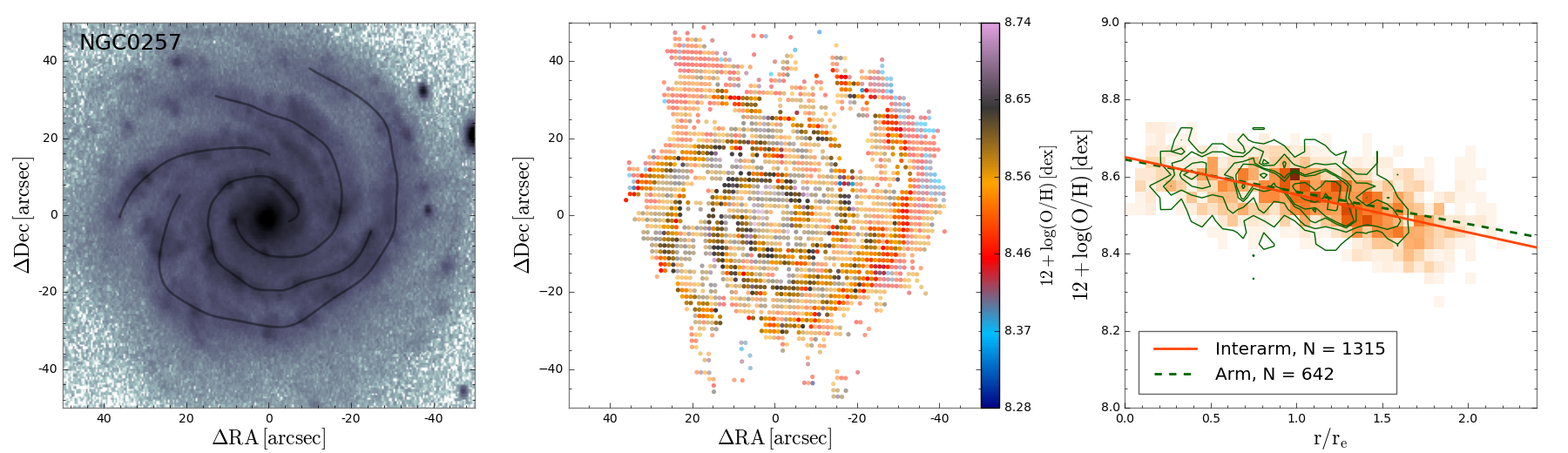}}
\caption{NGC0257. {\it Left panel:} Outline of the spiral arms on the deprojected SDSS $g$-band map. {\it Middle panel:} Colour map of the oxygen abundance distribution of the spiral arms (opaque dots) and the interarm region (transparent dots). {\it Right panel:} Radial density distribution in the oxygen abundance-galactocentric distance space of the spaxels located in the spiral arms (green contours) and those associated with the interarm region (orange colour map). The outermost contour encircles 100\% of the total number of spaxels, decreasing by 20\% in each consecutive contour. The lines represent the error-weighted linear fit derived for the arm (green dashed line) and interarm (orange solid line) distributions. The inset indicates the number of spaxels of each distribution that contribute to the derivation of the gradient, i.e. located within the radial range 0.5-2.0 $r_e$.}
\label{fig:NGC0257_mosaic}
\end{figure*}

The disc of spiral galaxies is known to be an axisymmetric component which sometimes appears elliptical, due to inclination effects. The greater the ellipticity, the more inclined the disc is with respect to our line of sight. Although we have selected galaxies with $i < 60 \degree$, the difference with respect to a face-on system is still appreciable and it can hamper the correct tracking of the spiral pattern in these discs. Thus, we need to deproject the galaxy discs to carry out the best possible characterisation of the spiral arms.

We deproject the galaxy images using the position angle (PA) and the inclination angle ($i$) derived as a by-product of the 2D photometric decomposition of the galaxies performed using the GASP2D algorithm. The actual implementation and a detailed description of the code is provided in \citet{mendezabreu2008,mendezabreu2014}. By performing the 2D decomposition, GASP2D provides us with the best combination of morphological components to fit the 2D observed light distribution of the analysed galaxies (bulge, disc, bar component, etc.). The results of the photometric decomposition of the entire CALIFA observed sample and the corresponding structural parameters is presented in \citet{mendezabreu2016a}. 

We must note that we have preferred not to correct for inclination effects in galaxies with inclinations below $35 \degr$ as the uncertainties on the derived correction exceed the very small effect on the spatial distribution of the spaxels.

\subsubsection{Arms classification}\label{sec:analysis2a}

The basis of a classification system for spiral galaxies according to their spiral arm structure was first introduced in \citet{elmegreen1982} and afterwards revised in \citet{elmegreen1987}. This way, spirals can be divided into flocculent or grand design galaxies according to the symmetry and continuity of the spiral pattern. Flocculent galaxies present small patchy spiral arms, while grand designs are characterised by the presence of long, symmetric, continuous arms. Initially, \citet{elmegreen1982} proposed 12 arm classes (AC) that were afterwards reduced to 10 categories in \citet{elmegreen1987} paying attention just to the spiral arm features and rejecting any distinction referring to the presence of bars or companion galaxies \citep[AC~$10-11$ according to][]{elmegreen1982}. Thus, galaxies with AC~$1-4$ are considered flocculent, and those with AC~$5-12$ (excluding AC 10 and 11) are grand design. Pure flocculent (AC $1-2$) and grand design galaxies (AC $12$) represent the opposite ends of the range and the intermediate classes show characteristics of both types, i.e. one prominent arm with other fragmented ones, tightly wrapped ringlike structures, or inner/outer symmetric arms together with feathery or irregular spiral patterns. The distinctions between some of these divisions is not straightforward as some galaxies can present characteristics of more than one class, being these cases very difficult to classify. For this reason, we decided to apply the general classification of flocculent/grand design galaxies without distinguishing between the different arm classes.

The classification was carried out independently by 14 authors of this work based on a visual inspection of the SDSS $g+u$ deprojected images\footnote{Using the seventh data release \citep[DR7,][]{dr7}.}, where the spiral arms are more easily outlined. We considered a classification to be valid when at least eight of the authors agree on the result, in all other cases the classification was considered ambiguous and the galaxy was discarded from further analysis. This way, the sample was reduced from 68 to 63 (discarding IC~1256, NGC~0477, NGC~1093, NGC~7321 and UGC~09476), of which 45 galaxies are classified as flocculent and 18 as grand design. Table~\ref{table} of Appendix~\ref{sec:appendix1} shows the final classification for all galaxies in the sample and Fig.~\ref{fig:ac_mosaic} presents a set of example images corresponding to flocculent (F) and grand  design galaxies (G). 

\subsubsection{Tracking of spiral arms}\label{sec:analysis2b}

We depict the spiral arms by visually tracing them on the SDSS $g$-band deprojected images to distinguish between the spaxels belonging to the spiral arms and the interarm regions. Then we interpolate the points using a cubic spline (a numeric function that is piecewise-defined by a third-degree polynomial). In the cases where the spiral arms are not continuous, the distinguishable fragmented parts are outlined. The left panel of Fig.~\ref{fig:NGC0257_mosaic} shows the spline fit of the detected spiral arms (black lines) superimposed on the SDSS image of NGC~0257. Once the spiral arms are traced, we assign them a width of 4 arcsec (see Sect.~\ref{sec:discussion} for an explanation of the possible effects on the results obtained using different arm width values), considering as spaxels belonging to the spiral arms those that are at a distance from the closest point of the arm (cubic spline function) of less than 2 arcsec. Thus, we consider that the remaining spaxels are associated with the interarm regions. The middle panel of Fig.~\ref{fig:NGC0257_mosaic} shows the distinction of the arm and interarm spaxels in an oxygen abundance distribution colour map, respectively represented as opaque and transparent dots.

This procedure has been used in previous works and has been proven successful in tracing other morphological features such as dust lanes in galactic bars \citep[see][]{sanchezmenguiano2015}.

\begin{figure*}
\begin{center}
\resizebox{0.95\hsize}{!}{\includegraphics{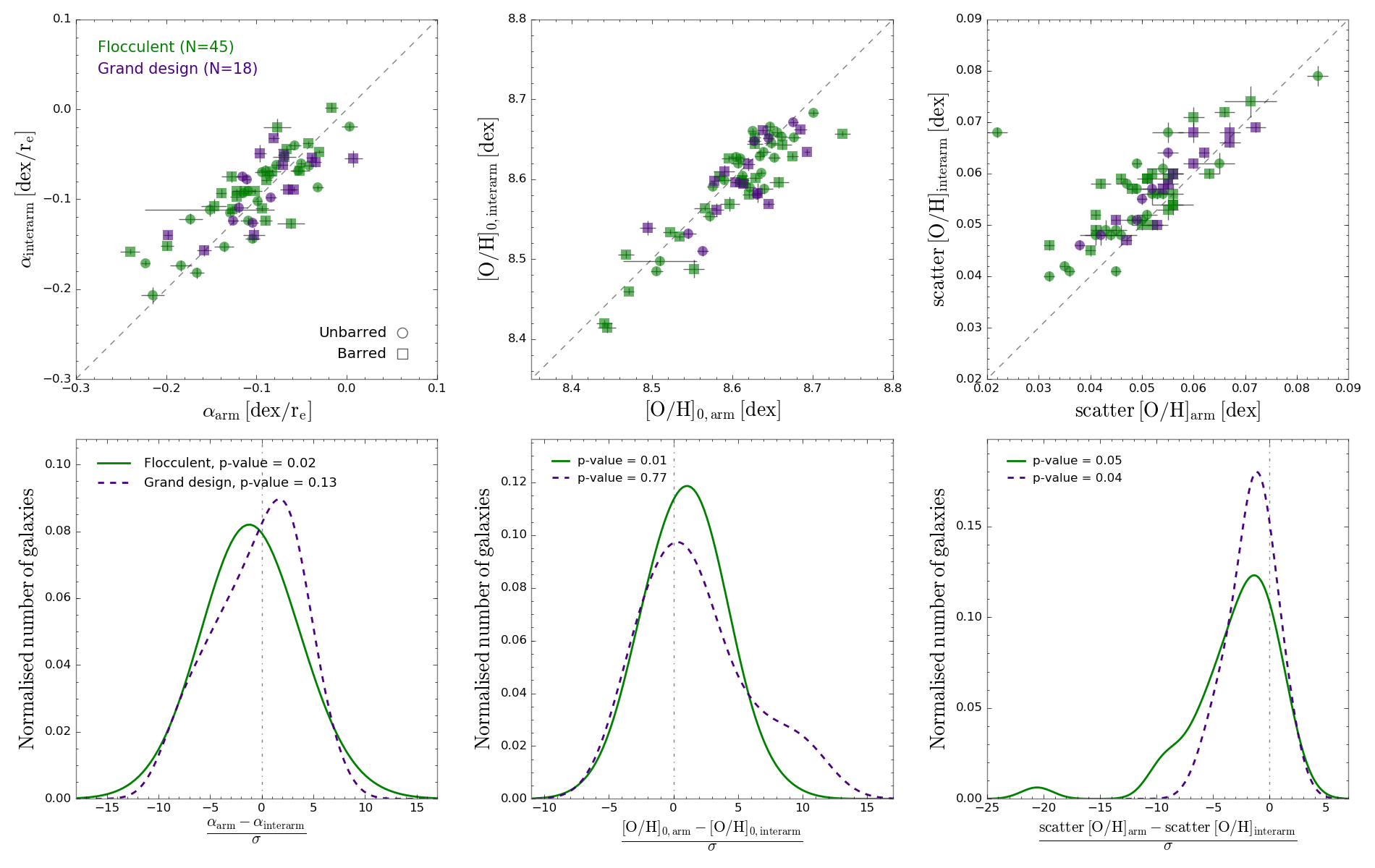}}
\caption{{\it Top panels:} Relation between the slope values (left), zero-points (middle) and scatter (right) of the radial linear fit performed considering arm gas abundance values (x-axes) and interarm values (y-axes). Flocculent galaxies are represented by green markers and grand design galaxies by purple markers. Squares and circles indicate barred and unbarred galaxies, respectively. The dashed grey lines represent the one-to-one relation. {\it Bottom panels:} Normalised distributions of the differences between the slope values (left), zero-points (middle) and scatter (right) of the interarm and arm regions divided by the quadratic sum of the errors derived for each galaxy. Green solid histograms represent the flocculent distributions and purple dashed histograms represent the distributions for the grand designs. The dashed dotted grey lines correspond to the zero value, indicating no differences between arm and interarm regions. The results of the t-tests are shown in the top left corners of the panels (see text for details).}
\label{fig:dif_gradients}
\end{center}
\end{figure*}

\subsection{Arm and interarm oxygen abundance gradients}\label{sec:analysis2c}

To derive the radial gradients of the oxygen abundance for the arm and the interarm regions of each galaxy we have normalised the galactocentric distance for the spaxels to the disc effective radius ($r_e$, see Appendix~A of \citealt{sanchez2014} for details on the computation of this parameter). Then we performed an error-weighted linear fit to the derived oxygen abundance values (see Sect.~\ref{sec:analysis1c}) between 0.5 and 2.0 $r_e$. We excluded the inner- and outermost regions, due to trends that deviate from the pure oxygen decrease (an inner drop and an outer flattening, see SM16 for more information). In general, for each galaxy we have approximately $400$ arm SF regions and $1200$ interarm SF regions that contribute to the derivation of the corresponding gradients, i.e. that are located in the radial range between 0.5 and 2.0 $r_e$.

In the right panel of Fig.~\ref{fig:NGC0257_mosaic} we present an example of the abundance profiles of the arm (green dashed line) and interarm regions (orange solid line) derived for the galaxy NGC~0257. The radial oxygen distribution for interarm regions is shown as an orange colour map and the distribution for the spiral arms is shown as a green contour map.


\section{Results}\label{sec:results}

In this paper we derive the oxygen abundance gradient separately for the SF regions belonging to the spiral arms and the interarm area in a sample of 63 galaxies using CALIFA data. The main goal of this analysis is to find possible differences in the chemical composition of the ionised gas associated with both components in order to shed some light on the origin of the spiral structure. We distinguish between flocculent and grand design galaxies as the differences observed in the spiral pattern may be the result of different formation processes of the spiral structure. 

\begin{figure*}
\begin{center}
\resizebox{0.8\hsize}{!}{\includegraphics{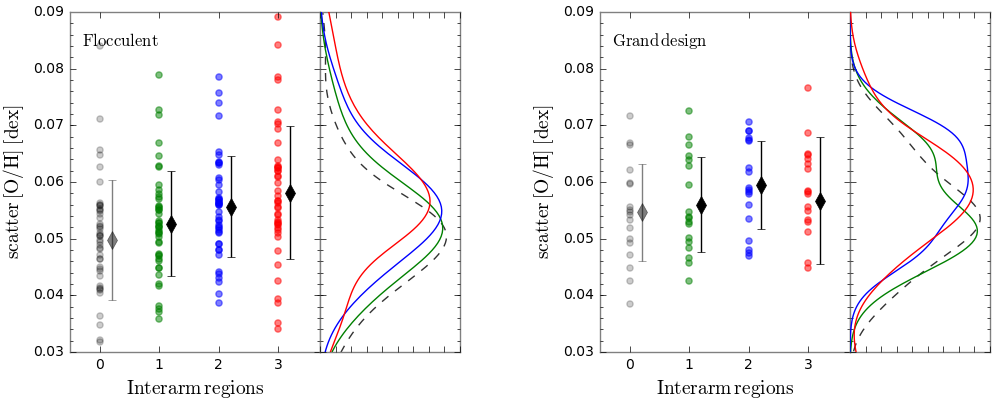}}
\caption{Distribution of the scatter of the oxygen abundances in the interarm regions according to the angular distance to the spiral arms for flocculent (left panel) and grand design (right panel) galaxies. Region 0 (grey) represents the arm region. Region 1 (green) is the interarm region closest to the spiral arms and region 3 (red) is the farthest away. Region 2 (blue) represents an intermediate region between 1 and 3. Each point represents the scatter of the SF spaxels within the corresponding interarm region for a particular galaxy. The black diamonds represent the mean values within each region; the error bars indicate the standard deviations. We also show the distribution of values for each region in the right auxiliary panels.}
\label{fig:scatter}
\end{center}
\end{figure*}

Figure~\ref{fig:dif_gradients} shows the outcome of the comparison of the derived gradients for the arm and interarm regions for all the individual galaxies. The top panels show the relation between the slope values (left), the zero-points (middle), and scatter (right) of the linear fits. The scatter is measured as the standard deviation of the differences between the observed abundances and the corresponding values from the linear regression according to their deprojected galactocentric distances. These parameters are also provided in Table~\ref{table}. Flocculent and grand design galaxies are represented by green and purple markers, respectively. We can see that for both distributions the values are slightly shifted from the one-to-one relation represented by the dashed grey lines, especially in the case of the dispersion. These deviations are observed mostly for the flocculent galaxies. However, we note the lower number of grand design systems (18) compared to the flocculent ones (45), which prevents us from reaching any strong conclusion about these types of spiral galaxies.

To show these differences more clearly, we represent in the bottom panels of Fig.~\ref{fig:dif_gradients} the distributions of the differences in slope values (left), zero-points (middle), and dispersions (right) between the arm and interarm regions divided by the quadratic sum of the errors. The shown distributions are not smoothed. They correspond to a density estimator in which each point is represented by a Gaussian distribution, centred in the actual value of the point, and with a sigma of the maximum distance between nearest neighbours. In this way by summing all the individual Gaussians we reproduce a smooth distribution (normalised to one) which is more peaky in the clustered points. The modal value of the differences between arm and interarm found for flocculent (grand design) galaxies are $\sim0.013$ dex/r$_e$ ($\sim0.015$ dex/r$_e$) for the slope, $\sim0.011$ dex ($\sim0.003$ dex) for the zero-point, and $\sim0.003$ dex ($\sim0.003$ dex) for the scatter in the linear fit. As can be inferred from these numbers, the observed differences are small. However, all of them are above 1-$\sigma$ (close to 2-$\sigma$ for some of them, see bottom panels of Fig.~\ref{fig:dif_gradients}), indicating that these differences are not due to the uncertainties in the derivation of the parameters.

We performed a Student's $t$-test for each of the six distributions to analyse whether these differences between the arm and interarm abundance distributions are statistically significant. In these tests, we compare the observed distributions (position of the peak and width) with a hypothetical Gaussian centred at zero (dashed vertical lines) and with a width of unity which represents the case where no differences are found between the arm and the interarm regions. The resulting p-values of these statistical tests are shown in the top left corner of the bottom panels of the figure. For flocculent galaxies, these p-values are below the significance level of 5\% (2\% in the slope, 1\% in the zero-point and exactly 5\% in the scatter), confirming that the observed differences are statistically significant. In the case of the grand design galaxies, the \mbox{p-values} are above the significance level for the slope and zero-point comparison (13\% and 77\% respectively) and below it for the scatter (4\%). This indicates that only the differences in the scatter with respect to the linear fit are statistically significant for these galaxies. However, again, we must be careful with the conclusions we obtain for the analysis of the grand design galaxies because of the low number statistics of these systems in the sample. The inclusion of more grand designs is needed to confirm the results of the Student's $t$-test on these galaxies.

The right panels of Fig.~\ref{fig:dif_gradients} show that in general the scatter of the oxygen abundances is higher in the interarm regions than within the spiral arms for both flocculent and grand design galaxies. The 2D coverage of the CALIFA data allows us to assess whether this scatter is the same in the whole interarm region or if there is a dependency on the angular distance from the arms. With this purpose we separate the spaxels of the interarm regions in three groups according to the distance to the closest arm, region 1 being the closest to the spirals and region 3 the farthest away. Region 2 is an intermediate region between 1 and 3. Using only the spaxels belonging to each of these regions we derive the scatter with respect to the linear regression in a similar way as done for the entire interarm region (top right panel of Fig.~\ref{fig:dif_gradients}). The outcome of this test is shown in Fig.~\ref{fig:scatter} for flocculent (left panel) and grand design (right panel) galaxies. Each point represents the scatter of the SF regions within the corresponding interarm region (green for region 1, blue for region 2, and red for region 3) for a particular galaxy. The scatter for the SF regions within the spiral arms is also represented for a better comparison (grey markers). The black diamonds represent the mean values within each region; the error bars indicate the standard deviations. The distribution of values for each region is represented in the right auxiliary panels. For flocculent galaxies, we see that the mean value of the scatter (black diamond) increases when moving away from the spiral arms. However, for grand design galaxies the scatter is larger when moving from region 1 to 2, but if we get a bit farther away (region 3) then it decreases again, indicating that there is no trend of increasing the scatter when moving away from the arms. 

\begin{figure*}
\begin{center}
\resizebox{0.95\hsize}{!}{\includegraphics{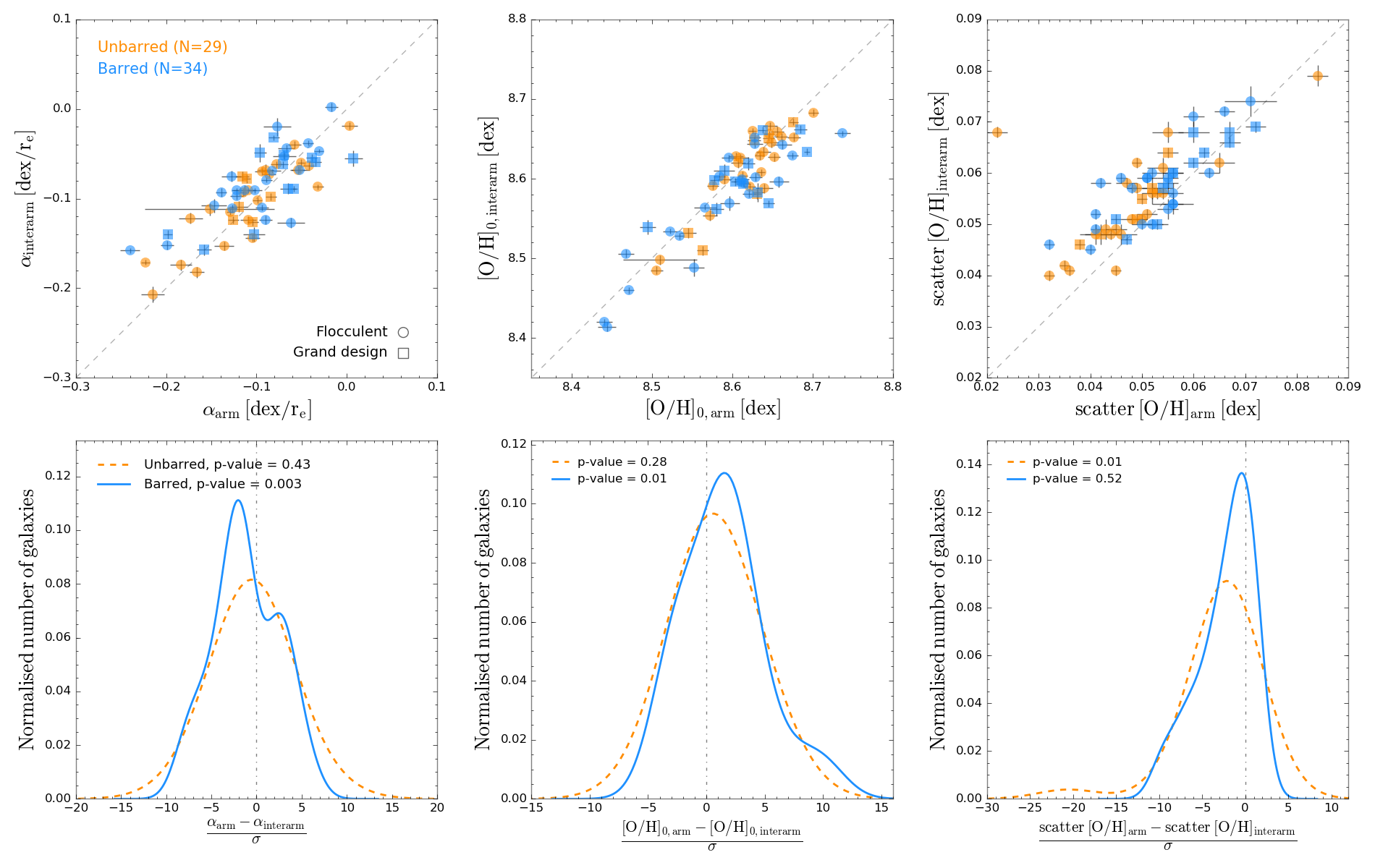}}
\caption{Same as Fig.~\ref{fig:dif_gradients} but distinguishing between barred (blue markers and  solid histograms) and unbarred (orange markers and dashed histograms) galaxies. See caption above for more details.}
\label{fig:dif_bars}
\end{center}
\end{figure*}

Apart from the comparison between flocculent and grand design galaxies, we performed the same tests dividing the sample into barred and unbarred galaxies. This comparison is shown in Fig.~\ref{fig:dif_bars}; as in the previous case, the barred and unbarred galaxies are represented by blue and orange colours, respectively. In this case we have a similar number of barred (34) and unbarred (29) galaxies. The averaged absolute differences found for unbarred (barred) galaxies are $\sim0.005$ dex/r$_e$ ($\sim0.020$ dex/r$_e$) for the slope, $\sim0.006$ dex ($\sim0.016$ dex) for the zero-point, and $\sim0.005$ dex ($\sim0.001$ dex) for the scatter in the linear fit. The performed t-tests yield \mbox{p-values} for unbarred galaxies that are clearly above the significance level of 5\% in the slope and the zero-point (around 40\% and 30\%, respectively), but below it in the scatter (1\%), indicating that the only significant differences are found between the arm and interarm abundance gradients regarding the scatter for these unbarred systems. On the other hand, small differences are obtained for the barred galaxies (\mbox{p-values} of 0.3\% and 1\% respectively) in the slope and zero-point of the gradients. No statistically significant differences are found in the scatter of the oxygen gradient between the arm and interarm distributions for the barred galaxies. 


\section{Discussion}\label{sec:discussion}

In this work we have carried out a study of the arm and interarm abundance distributions in a sample of 63 CALIFA spiral galaxies performing a spaxel-by-spaxel analysis. We distinguish between flocculent and grand design galaxies in order to better understand the origin of the spiral structure. For this purpose, we have derived oxygen abundance gradients for both distributions (arm and interarm), to compare the characteristic parameters of the profiles (slope, zero-point, and dispersion) between them. We have also distinguished between barred and unbarred galaxies to assess the effect of the presence of bars in the abundance distribution.

The derived results in this study rely strongly on a good arm classification of the galaxies. For this reason, in order to confirm the strength of the results, we have repeated the previously described analysis considering this time only the galaxies for which at least $75\%$ of the authors (10 out of 14) agree on the spiral classification. This way, with a reduced sample of 41 galaxies (30 flocculent and 11 grand design) we obtain qualitatively equivalent results.

Another aspect of this analysis that could affect the results is the distinction between the arm and the interarm regions. As stated in Sect.~\ref{sec:analysis2b}, we selected a width of 4 arcsec to derive the arm abundance distribution. This value was chosen by visually examining the SDSS $g$-band images of the galaxies. As all the galaxies in the sample are located at a similar redshift, we selected an average width of the arm (measured in arcsec) that corresponds approximately to the same physical width for all of them ($\sim1.5$~kpc). We checked whether using the same value for the arm width along the arm extension and for all the galaxies is a good approximation. To this end, we made a perpendicular cut to the arm at three different positions on the $g$-band images of the galaxies. At the position of the arm centre the flux profile presents a bump, due to the presence of the arm. Fitting a Gaussian and measuring the FWHM we can estimate the arm width. Performing this test for some galaxies with well-defined spiral arms, we find that there is neither a significant dependence of the arm width on radius nor between galaxies. We find that on average, a 4 arcsec-width is representative of this sample. We also performed several tests regarding the effect of using other values of the arm width in the analysis. If we use a width of 2 arcsec we obtain similar results. As we increase this value, the observed differences begin to disappear. When we reach a value of 8 arcsec, these differences become statistically insignificant according to the $t$-tests. This result is expected; by increasing the arm width, we include more SF regions belonging to the interarm regions in the analysis and the differences are diluted.

As we mentioned in Sect~\ref{sec:analysis1b}, the results may also be affected by the required S/N of the emission lines involved in the oxygen derivation. In \citet{sanchezmenguiano2016} we already showed that a 1-$\sigma$ cut is enough to reproduce the radial trends of the gas-phase oxygen abundance from \hii\,regions. However, considering that in this work we are going a step further by comparing the abundance distribution between the arm and interarm regions, we also assess how this cut affects our results. Applying a 3-$\sigma$ cut reduces the number of galaxies for which we cover a substantial radial range to accurately derive the abundance gradient. Despite the loss of eight flocculent galaxies and the general decrease in the number of the spaxels analysed (especially for the interarm region) in each galaxy, the results point to the same conclusions. The most significant difference is seen when analysing the scatter around the abundance gradient for the grand design galaxies, which now present a similar scatter in the arm and the interarm regions. However, this is somewhat expected since we have reduced the number of spaxels located in the interarm regions. This effect is especially significant in the grand design galaxies where the gas is more concentrated in the spiral arms. The agreement we find supports our methodology and strengthens our conclusions, and therefore we consider that the adopted flux cut is a valid selection criterion.

The arm abundance distribution for each galaxy is determined considering the spaxels belonging to all the traced spiral arms together. This increases the number statistics of the arm region, and significantly improves the radial coverage, which in some cases is very limited for individual arms. However, by considering all the spaxels associated with the spiral arms together we are assuming that the physical conditions of the \hii\,regions belonging to the spiral arms do not depend on their association with a particular arm (if no external factors such as a recent merging that can produce differences in the abundance distribution are involved), which is not a strong assumption since the spiral arms are supposed to form a single structure with a common origin. A similar reasoning leads us to group all the spaxels not belonging to the spiral arms together, which we call the interarm region.

It is beyond the purpose of this work to make a detailed study of the effects of using different empirical calibrators in the analysis. However, to confirm our results, we have also derived the oxygen radial distributions using some of the most commonly used empirical calibrators: the \citet{pilyugin2010} calibration for the ONS index and the \citet{dopita2013} calibration based on the MAPPINGS IV code developed by the authors (called `pyqz'). Qualitatively equivalent results have been found when using the pyqz calibrator. However, in the case of the ONS indicator, only the results regarding the scatter relative to the gradient, which are the largest differences found in the analysis, are compatible. The differences in the slope and zero-point values, which are of the order of 2-$\sigma$, are diluted owing to the larger errors associated with the ONS calibrator (approximately three times the errors derived with O3N2). Calibrators based on R23 like ONS are affected by well-known problems (non-linear dependence on O/H; larger errors associated with \mbox{[\oii]~$\lambda3727$} due to the spectrophotometric calibration of CALIFA data, \citealt{sanchez2012a}; need to correct for dust attenuation due to the distance in wavelength between the involved lines, etc.); instead, calibrators based on O3N2 index are not affected by these factors, which makes the \citet{marino2013} calibrator able to distinguish between the arm and interarm abundance distributions with greater precision.

The results of our analysis show that subtle differences exist between the arm and interarm radial abundance distribution. After performing Student's $t$-tests we conclude that only the differences found in flocculent and barred galaxies are statistically significant. On average, we find that the ionised gas in the interarm regions of these subgroups of galaxies presents a shallower gradient with a lower zero-point value and a larger dispersion in the oxygen abundances. Furthermore, we find that for flocculent galaxies this dispersion correlates with the angular distance to the spiral arms: the farthest spaxels present the larger scatter. Grand design galaxies, on the other hand, do not seem to present any trend in the dispersion with the distance to the arms. The extension of this study to a larger sample of galaxies (with better spatial resolution) is needed to confirm whether the small differences found for flocculent and barred systems are real or just due to uncertainties in the determination of the abundances. This extension is also necessary for grand design systems due to the low number statistics of these objects analysed in this work. Moreover, we have to be aware of the limitations of the conclusions arising from the results of Student's $t$-tests, as these kinds of statistical tests always assume an underlying Gaussian distribution for the parameters, which may sometimes not be the case.

Our work does not contradict the results of \citet{zinchenko2016}, who also analyse galaxies from the CALIFA survey. We note that the observed differences are subtle and limited to small areas of the discs, the spiral arms. These differences between the arm and the interarm regions would be diluted in a study of the asymmetries of the overall abundance distribution such as the one performed by \citet{zinchenko2016}.

The largest differences have been detected when analysing the scatter with respect to the negative gradient. We find that this dispersion is larger in the interarm regions than along the spiral arms. We checked that these results do not rely on the use of a larger number of spaxels belonging to the interarm regions. We randomly selected from the spaxels of the interarm area a subsample of equal size to the arm area, and performed the derivation of the dispersion with this subset of spaxels. The conclusion is the same: the interarm regions present higher scatter in the abundance values than the arm regions with respect to the negative gradient. This is a strong indicator that spirals are important drivers of mixing in galactic discs via the gas shocks that they induce, and maintain a nearly homogeneous chemistry along the streamlines of the gas. However interarm gas is polluted stochastically by young stars forming in situ, leading to an increased scatter.

Regarding the comparison of flocculent and grand design galaxies, our results agree with the current interpretation of spiral structure theories, according to which the spiral arms of flocculent galaxies are generated by density variations driven by local instabilities, while in grand design galaxies the spiral arms are caused by quasi-stationary density waves \citep{dobbs2014}. The fact that grand design galaxies do not exhibit statistically significant differences in the abundance distribution between the arm and interarm regions (see Fig.~\ref{fig:dif_gradients}) is in agreement with the presence of density waves that affect the gas content of the entire galaxy as these features move across the disc. In addition, the fact that the scatter in the oxygen abundances does not depend on the distance to the spiral arms for grand design systems supports this idea (see Fig.~\ref{fig:scatter}). The regions closest to the spiral arms change as the density waves move across the galaxy discs and this makes the scatter similar in the whole interarm region. On the other hand, in flocculent galaxies, density variations produced by local instabilities only affect specific locations of the discs associated with the spiral arms, making the gas content of these regions (and therefore the chemical composition) different from that found in the interarm regions. The fact that the differences found in this work for these galaxies are small can be explained taking into account that these local instabilities are not static, but are continuously formed and destroyed across the disc, not affecting the same material over a long period of time. In this case, during the lifetime of the spiral arms, the closest areas to them are always the same and, because they are affected by the presence of the arms, due to their proximity, the scatter is lower (as occurs within the spiral arms but to a lesser extent). As we move away from the arms, the gas is less affected by the mixing of the arms and the scatter increases.

Although our results can be explained by this difference in the origin of the spiral structure, other alternative interpretations are also possible. In fact, it is plausible that both flocculent and grand design galaxies are the result of the same mechanism generating the spiral structure. In that case, the differences found in the abundance distribution between the arm and interarm regions in flocculent galaxies (see Fig.~\ref{fig:dif_gradients}) can be explained simply because the gas outside the spirals is stochastically polluted more than in the grand design galaxies where the star formation is more concentrated in the spiral arms. This could also be the reason why the metallicity scatter increases with increasing distance from the spiral in flocculent galaxies, whereas no statistically significant difference in the scatter of interarm gas by location is found for grand design galaxies (Fig.~\ref{fig:scatter}). Another explanation for the differences found between flocculent and grand design galaxies is that in grand design spirals the gas mixing could be more efficient. However, we find that the scatter in grand design galaxies is on average higher than in flocculent galaxies (in the arm and in the interarm regions, see Fig.~\ref{fig:scatter}), allowing us to rule out this interpretation. 

The analysis of barred and unbarred galaxies shows that only barred galaxies present (small) differences in the radial oxygen abundance distribution between the arm and interarm regions. We have to be cautious with this conclusion as these differences are even more subtle than those found for flocculent galaxies. Again, an extended study with more galaxies (and better spatial resolution) is necessary to confirm these preliminary results. Despite these caveats, this result may suggest that the presence of a bar induces differences in the abundance distribution of the spiral galaxies. Bars have been proposed as a key mechanism in the dynamical evolution of disc galaxies. For instance, recent numerical simulations from \citet{dimatteo2013} have shown that radial migration induced by a bar leads to significant azimuthal variations in the metallicity distribution of old stars. More recently, \citet{sanchezmenguiano2016b} showed that radial migration can also affect the gas metallicity distribution producing differences in the abundance distribution of the spiral arms; however, these differences do not seem to affect the overall abundance gradient for the gas \citep{sanchez2014,marino2016,sanchezmenguiano2016} or for the stars \citep{sanchezblazquez2014}.

In light of these results, we suggest that bars and flocculent arms are two mechanisms producing differences in the abundance distribution between arm and interarm regions. In order to check the robustness of this statement, we divide the flocculent galaxies according to the presence or absence of bars. If these mechanisms are independent, then the subgroup of flocculent barred galaxies (where both mechanisms come into play) is expected to present the highest differences. Indeed, when performing this test, we obtain lower p-values for this subset of galaxies (1\% versus 52\% in the slope distributions and 1\% versus 30\% in the zero-point distributions for flocculent barred and flocculent unbarred galaxies, respectively). Owing to the low number of grand design galaxies, we could not carry out the same test for these systems.

\section{Conclusions}\label{sec:conclusions}

The existence of a radial decrease in the gas chemical abundances of spiral galaxies was well established by observations decades ago \citep{searle1971, martin1992, kennicutt2003, pilyugin2004, rosalesortega2011, bresolin2012, marino2012, sanchez2012b, sanchez2014, sanchezmenguiano2016}, supporting the inside-out scenario for disc evolution. However, only a few works have studied possible azimuthal differences in the derived gradient associated with the presence of the spiral structure, which is still an open question.

To our knowledge, this is the first observational work finding differences between the gas abundance of spiral arms and interarm regions using a large sample of galaxies. Our analysis yields differences that are subtle and statistically significant only for flocculent galaxies. This result suggests that the mechanisms generating the spiral structure in these galaxies may be different to those producing grand design arms. Another possibility is that these differences may be due to a higher star formation {\it outside} the spiral arms for the flocculent galaxies and a more concentrated star formation in the spirals for the grand design ones. The small differences between the arm and the interarm abundance distributions in barred galaxies suggests that bars may have a direct effect on the chemical distribution of these galaxies, even though these differences are not reflected in the overall observed abundance distribution.

The extension of this analysis to a larger sample of galaxies with data of higher quality in terms of spatial resolution and S/N per spatial unit is needed to confirm these results. As a preliminary result, this work finds that bars and flocculent arms are two independent mechanisms that generate differences in the abundance distribution between the spiral arms and the interarm regions. So far, this is the first observational work finding such differences in isolated galaxies.

\vspace{0.5cm}
\begin{acknowledgements}
This study makes use of the data provided by the Calar Alto Legacy Integral Field Area (CALIFA) survey (\url{http://califa.caha.es/}) based on observations collected at the Centro Astron\'omico Hispano Alem\'an (CAHA) at Calar Alto, operated jointly by the Max-Planck Institut f\"ur Astronomie and the Instituto de Astrof\'isica de Andaluc\'ia (CSIC).\\

CALIFA is the first legacy survey being performed at Calar Alto. The CALIFA collaboration would like to thank the IAA-CSIC and MPIA-MPG as major partners of the observatory, and CAHA itself, for the unique access to telescope time and support in manpower and infrastructures. The CALIFA collaboration also thanks the CAHA staff for the dedication to this project.\\

We would like to thank the anonymous referee for comments which helped to improve the content of this paper and the confidence of our results.\\

We acknowledge financial support from the Spanish {\em Ministerio de Econom\'ia y Competitividad (MINECO)} via grants AYA2012-31935 and AYA2014-53506-P, and from the ``Junta de Andaluc\'ia'' local government through the FQM-108 project. We also acknowledge support from the ConaCyt funding program 180125 and DGAPA IA100815. V.P.D. is supported by STFC Consolidated grant \#ST/M000877/1. V.P.D. acknowledges being a part of the network supported by the COST Action TD1403 ``Big Data Era in Sky and Earth Observation''. L.G. was supported in part by the US National Science Foundation under Grant AST-1311862. R.A.M. acknowledges support from the Swiss National Science Foundation. J.M.A. acknowledges support from the European Research Council Starting Grant (SEDmorph; P.I. V. Wild) and MINECO through the grant AYA2013-43188-P. I.M. acknowledges support from the Junta de Andalucia through project TIC114, and the MINECO through projects AYA2013-42227-P and AYA2016-76682C3-1-P. Y.A. is financially supported by the \emph{Ram\'{o}n y Cajal} programme (contract RyC-2011-09461) and project AYA2016-79724-C4-1-P from the Spanish MINECO, as well as the exchange programme SELGIFS FP7-PEOPLE-2013-IRSES-612701 funded by the EU. R.M.G.D. acknowledges support from the Spanish grant AYA2010-15081, and from the ``Junta de Andaluc\'ia'' FQ1580 project.\\

This research makes use of python (\url{http://www.python.org}); Matplotlib \citep[][]{hunter2007}, a suite of open-source python modules that provide a framework for creating scientific plots; and Astropy, a community-developed core Python package for Astronomy \citep[][]{astropy2013}.

\end{acknowledgements}

\bibliographystyle{aa} 
\bibliography{bibliography}

\clearpage
\onecolumn

\appendix
\section{Galaxy classification and oxygen abundance information}\label{sec:appendix1} 

In this section we present a table with information derived in this work related to the arm and bar/unbarred classifications, and the radial oxygen abundance distribution for all the galaxies in the sample. From left to right the columns correspond to

\begin{enumerate}
\item galaxy name;
\item barred (B) or unbarred (U) galaxy, derived from GASP2D photometric decomposition (see Sect.~\ref{sec:analysis2});
\item flocculent (F) or grand design (G) galaxy;
\item slope of the oxygen abundance gradient of the arm region;
\item slope of the oxygen abundance gradient of the interarm region;
\item zero-point of the oxygen abundance gradient of the arm region;
\item zero-point of the oxygen abundance gradient of the interarm region;
\item dispersion of the oxygen abundance gradient of the arm region;
\item dispersion of the oxygen abundance gradient of the interarm region.
\end{enumerate}
The errors on the measured parameters are given in parentheses.

\centering
\begin{longtab}
\tabcolsep=0.2cm
\LTcapwidth=\textwidth
\begin{longtable}{l@{\hspace{0.9cm}}ccccccccc@{\hspace{-0.2cm}}}
\caption{Arm classification and oxygen abundance information. Details are given in Appendix~\ref{sec:appendix1} above.}\\
\hline\hline\\
Name & B/U & F/G & $\rm \alpha_{ar}$ & $\rm \alpha_{in}$ & $\rm [O/H]_{0,ar}$ & $\rm [O/H]_{0,in}$ & $\rm  \Delta[O/H]_{ar}$ & $\rm \Delta[O/H]_{in}$\\[0.1cm]
  &  &  & [dex/$r_e$] & [dex/$r_e$] & [dex] & [dex]  & [dex] & [dex] \\[0.2cm]
{\tiny(a)} & {\tiny(b)} & {\tiny(c)} & {\tiny(d)} & {\tiny(e)} & {\tiny(f)} & {\tiny(g)} & {\tiny(h)} & {\tiny(i)} & \\[0.1cm]
\hline\\
\endfirsthead
\caption{Continued.}\\
\hline\hline\\
Name & B/U & F/G & $\rm \alpha_{ar}$ & $\rm \alpha_{in}$ & $\rm [O/H]_{0,ar}$ & $\rm [O/H]_{0,in}$ & $\rm \Delta[O/H]_{ar}$ & $\rm \Delta[O/H]_{in}$\\[0.1cm]
  &  &  & [dex/$r_e$] & [dex/$r_e$] & [dex] & [dex]  & [dex] & [dex] \\[0.2cm]
{\tiny(a)} & {\tiny(b)} & {\tiny(c)} & {\tiny(d)} & {\tiny(e)} & {\tiny(f)} & {\tiny(g)} & {\tiny(h)} & {\tiny(i)} \\[0.1cm]
\hline\\
\endhead
NGC0001  &  U  &  F  & $ -0.051 \,( 0.004 )$  & $ -0.060 \,( 0.004 )$  & $ 8.634 \,( 0.006 )$  & $ 8.629 \,( 0.005 )$  & $ 0.052 \,( 0.002 )$  & $ 0.056 \,( 0.002 )$ \\
NGC0036  &  B  &  G  & $ -0.065 \,( 0.009 )$  & $ -0.089 \,( 0.006 )$  & $ 8.620 \,( 0.008 )$  & $ 8.619 \,( 0.008 )$  & $ 0.055 \,( 0.002 )$  & $ 0.058 \,( 0.002 )$ \\
NGC0214  &  B  &  F  & $ -0.052 \,( 0.011 )$  & $ -0.068 \,( 0.003 )$  & $ 8.628 \,( 0.010 )$  & $ 8.644 \,( 0.005 )$  & $ 0.055 \,( 0.002 )$  & $ 0.053 \,( 0.002 )$ \\
NGC0234  &  U  &  F  & $ -0.032 \,( 0.006 )$  & $ -0.087 \,( 0.003 )$  & $ 8.625 \,( 0.005 )$  & $ 8.660 \,( 0.004 )$  & $ 0.049 \,( 0.002 )$  & $ 0.057 \,( 0.001 )$ \\
NGC0237  &  U  &  F  & $ -0.109 \,( 0.006 )$  & $ -0.091 \,( 0.003 )$  & $ 8.661 \,( 0.008 )$  & $ 8.653 \,( 0.004 )$  & $ 0.044 \,( 0.002 )$  & $ 0.048 \,( 0.001 )$ \\
NGC0257  &  U  &  G  & $ -0.084 \,( 0.006 )$  & $ -0.098 \,( 0.004 )$  & $ 8.645 \,( 0.007 )$  & $ 8.651 \,( 0.004 )$  & $ 0.052 \,( 0.002 )$  & $ 0.057 \,( 0.002 )$ \\
NGC0309  &  B  &  G  & $ -0.158 \,( 0.008 )$  & $ -0.157 \,( 0.006 )$  & $ 8.685 \,( 0.008 )$  & $ 8.662 \,( 0.005 )$  & $ 0.053 \,( 0.002 )$  & $ 0.050 \,( 0.001 )$ \\
NGC0496  &  U  &  F  & $ -0.136 \,( 0.011 )$  & $ -0.153 \,( 0.005 )$  & $ 8.610 \,( 0.010 )$  & $ 8.626 \,( 0.005 )$  & $ 0.054 \,( 0.003 )$  & $ 0.061 \,( 0.002 )$ \\
NGC0716  &  B  &  F  & $ -0.128 \,( 0.011 )$  & $ -0.075 \,( 0.006 )$  & $ 8.658 \,( 0.012 )$  & $ 8.596 \,( 0.007 )$  & $ 0.056 \,( 0.004 )$  & $ 0.054 \,( 0.002 )$ \\
NGC0768  &  U  &  G  & $ -0.116 \,( 0.008 )$  & $ -0.075 \,( 0.005 )$  & $ 8.563 \,( 0.007 )$  & $ 8.510 \,( 0.006 )$  & $ 0.055 \,( 0.002 )$  & $ 0.064 \,( 0.001 )$ \\
NGC0776  &  B  &  G  & $ -0.081 \,( 0.006 )$  & $ -0.032 \,( 0.003 )$  & $ 8.645 \,( 0.007 )$  & $ 8.569 \,( 0.004 )$  & $ 0.060 \,( 0.003 )$  & $ 0.068 \,( 0.002 )$ \\
NGC0873  &  U  &  F  & $ -0.078 \,( 0.006 )$  & $ -0.062 \,( 0.004 )$  & $ 8.613 \,( 0.006 )$  & $ 8.604 \,( 0.005 )$  & $ 0.032 \,( 0.001 )$  & $ 0.040 \,( 0.001 )$ \\
NGC0991  &  B  &  F  & $ -0.067 \,( 0.009 )$  & $ -0.044 \,( 0.004 )$  & $ 8.471 \,( 0.007 )$  & $ 8.460 \,( 0.004 )$  & $ 0.048 \,( 0.001 )$  & $ 0.057 \,( 0.001 )$ \\
NGC1094  &  U  &  G  & $ -0.111 \,( 0.004 )$  & $ -0.078 \,( 0.005 )$  & $ 8.631 \,( 0.005 )$  & $ 8.581 \,( 0.006 )$  & $ 0.042 \,( 0.004 )$  & $ 0.048 \,( 0.002 )$ \\
NGC1659  &  B  &  F  & $ -0.094 \,( 0.007 )$  & $ -0.110 \,( 0.003 )$  & $ 8.596 \,( 0.009 )$  & $ 8.626 \,( 0.004 )$  & $ 0.051 \,( 0.002 )$  & $ 0.059 \,( 0.001 )$ \\
NGC1667  &  B  &  F  & $ -0.031 \,( 0.005 )$  & $ -0.047 \,( 0.002 )$  & $ 8.584 \,( 0.007 )$  & $ 8.603 \,( 0.004 )$  & $ 0.040 \,( 0.001 )$  & $ 0.045 \,( 0.001 )$ \\
NGC2253  &  B  &  F  & $ -0.017 \,( 0.007 )$  & $ +0.002 \,( 0.005 )$  & $ 8.597 \,( 0.012 )$  & $ 8.569 \,( 0.009 )$  & $ 0.055 \,( 0.002 )$  & $ 0.059 \,( 0.002 )$ \\
NGC2347  &  B  &  F  & $ -0.122 \,( 0.006 )$  & $ -0.091 \,( 0.003 )$  & $ 8.675 \,( 0.007 )$  & $ 8.629 \,( 0.004 )$  & $ 0.063 \,( 0.002 )$  & $ 0.060 \,( 0.001 )$ \\
NGC2530  &  B  &  F  & $ -0.062 \,( 0.015 )$  & $ -0.127 \,( 0.006 )$  & $ 8.468 \,( 0.010 )$  & $ 8.506 \,( 0.005 )$  & $ 0.052 \,( 0.002 )$  & $ 0.060 \,( 0.001 )$ \\
NGC2540  &  B  &  F  & $ -0.122 \,( 0.007 )$  & $ -0.097 \,( 0.004 )$  & $ 8.611 \,( 0.009 )$  & $ 8.596 \,( 0.005 )$  & $ 0.042 \,( 0.002 )$  & $ 0.058 \,( 0.001 )$ \\
NGC2604  &  B  &  F  & $ -0.069 \,( 0.013 )$  & $ -0.053 \,( 0.004 )$  & $ 8.441 \,( 0.010 )$  & $ 8.420 \,( 0.004 )$  & $ 0.051 \,( 0.001 )$  & $ 0.059 \,( 0.001 )$ \\
NGC2906  &  U  &  F  & $ -0.058 \,( 0.007 )$  & $ -0.040 \,( 0.004 )$  & $ 8.652 \,( 0.008 )$  & $ 8.627 \,( 0.005 )$  & $ 0.035 \,( 0.001 )$  & $ 0.042 \,( 0.001 )$ \\
NGC2916  &  U  &  G  & $ -0.126 \,( 0.006 )$  & $ -0.124 \,( 0.004 )$  & $ 8.676 \,( 0.006 )$  & $ 8.671 \,( 0.004 )$  & $ 0.038 \,( 0.001 )$  & $ 0.046 \,( 0.001 )$ \\
NGC3057  &  B  &  F  & $ -0.147 \,( 0.014 )$  & $ -0.108 \,( 0.008 )$  & $ 8.444 \,( 0.011 )$  & $ 8.414 \,( 0.006 )$  & $ 0.066 \,( 0.002 )$  & $ 0.072 \,( 0.001 )$ \\
NGC3381  &  B  &  F  & $ -0.083 \,( 0.010 )$  & $ -0.069 \,( 0.002 )$  & $ 8.566 \,( 0.014 )$  & $ 8.564 \,( 0.003 )$  & $ 0.056 \,( 0.002 )$  & $ 0.054 \,( 0.001 )$ \\
NGC3614  &  U  &  F  & $ -0.215 \,( 0.013 )$  & $ -0.207 \,( 0.009 )$  & $ 8.649 \,( 0.009 )$  & $ 8.645 \,( 0.006 )$  & $ 0.051 \,( 0.002 )$  & $ 0.052 \,( 0.001 )$ \\
NGC3811  &  U  &  F  & $ -0.099 \,( 0.005 )$  & $ -0.102 \,( 0.003 )$  & $ 8.646 \,( 0.006 )$  & $ 8.656 \,( 0.004 )$  & $ 0.045 \,( 0.001 )$  & $ 0.041 \,( 0.001 )$ \\
NGC4047  &  U  &  F  & $ -0.129 \,( 0.004 )$  & $ -0.115 \,( 0.003 )$  & $ 8.701 \,( 0.006 )$  & $ 8.683 \,( 0.004 )$  & $ 0.045 \,( 0.002 )$  & $ 0.049 \,( 0.001 )$ \\
NGC4210  &  B  &  G  & $ -0.059 \,( 0.004 )$  & $ -0.089 \,( 0.003 )$  & $ 8.638 \,( 0.006 )$  & $ 8.661 \,( 0.004 )$  & $ 0.047 \,( 0.001 )$  & $ 0.047 \,( 0.001 )$ \\
NGC5000  &  B  &  G  & $ -0.096 \,( 0.007 )$  & $ -0.049 \,( 0.010 )$  & $ 8.632 \,( 0.007 )$  & $ 8.583 \,( 0.012 )$  & $ 0.067 \,( 0.002 )$  & $ 0.066 \,( 0.001 )$ \\
NGC5016  &  U  &  F  & $ -0.109 \,( 0.009 )$  & $ -0.124 \,( 0.004 )$  & $ 8.647 \,( 0.009 )$  & $ 8.666 \,( 0.005 )$  & $ 0.046 \,( 0.003 )$  & $ 0.048 \,( 0.002 )$ \\
NGC5056  &  B  &  F  & $ -0.089 \,( 0.006 )$  & $ -0.079 \,( 0.003 )$  & $ 8.523 \,( 0.009 )$  & $ 8.534 \,( 0.003 )$  & $ 0.046 \,( 0.001 )$  & $ 0.059 \,( 0.001 )$ \\
NGC5320  &  B  &  F  & $ -0.240 \,( 0.011 )$  & $ -0.158 \,( 0.003 )$  & $ 8.737 \,( 0.010 )$  & $ 8.657 \,( 0.003 )$  & $ 0.052 \,( 0.002 )$  & $ 0.050 \,( 0.001 )$ \\
NGC5376  &  U  &  F  & $ +0.003 \,( 0.009 )$  & $ -0.019 \,( 0.004 )$  & $ 8.607 \,( 0.009 )$  & $ 8.620 \,( 0.006 )$  & $ 0.048 \,( 0.002 )$  & $ 0.051 \,( 0.001 )$ \\
NGC5520  &  B  &  F  & $ -0.102 \,( 0.008 )$  & $ -0.091 \,( 0.003 )$  & $ 8.612 \,( 0.009 )$  & $ 8.599 \,( 0.004 )$  & $ 0.032 \,( 0.001 )$  & $ 0.046 \,( 0.001 )$ \\
NGC5622  &  U  &  G  & $ -0.104 \,( 0.006 )$  & $ -0.126 \,( 0.004 )$  & $ 8.627 \,( 0.008 )$  & $ 8.648 \,( 0.005 )$  & $ 0.050 \,( 0.001 )$  & $ 0.055 \,( 0.001 )$ \\
NGC5633  &  U  &  F  & $ -0.086 \,( 0.007 )$  & $ -0.073 \,( 0.003 )$  & $ 8.677 \,( 0.008 )$  & $ 8.652 \,( 0.005 )$  & $ 0.036 \,( 0.001 )$  & $ 0.041 \,( 0.001 )$ \\
NGC5732  &  U  &  F  & $ -0.184 \,( 0.012 )$  & $ -0.174 \,( 0.006 )$  & $ 8.639 \,( 0.009 )$  & $ 8.634 \,( 0.006 )$  & $ 0.047 \,( 0.002 )$  & $ 0.058 \,( 0.001 )$ \\
NGC5735  &  B  &  G  & $ -0.198 \,( 0.005 )$  & $ -0.140 \,( 0.005 )$  & $ 8.693 \,( 0.004 )$  & $ 8.634 \,( 0.004 )$  & $ 0.067 \,( 0.002 )$  & $ 0.068 \,( 0.002 )$ \\
NGC5772  &  U  &  F  & $ -0.042 \,( 0.006 )$  & $ -0.063 \,( 0.004 )$  & $ 8.576 \,( 0.007 )$  & $ 8.591 \,( 0.005 )$  & $ 0.065 \,( 0.003 )$  & $ 0.062 \,( 0.002 )$ \\
NGC5829  &  U  &  G  & $ -0.120 \,( 0.010 )$  & $ -0.109 \,( 0.006 )$  & $ 8.545 \,( 0.010 )$  & $ 8.532 \,( 0.006 )$  & $ 0.049 \,( 0.002 )$  & $ 0.051 \,( 0.001 )$ \\
NGC5947  &  B  &  F  & $ -0.127 \,( 0.005 )$  & $ -0.111 \,( 0.004 )$  & $ 8.615 \,( 0.006 )$  & $ 8.595 \,( 0.005 )$  & $ 0.041 \,( 0.001 )$  & $ 0.052 \,( 0.001 )$ \\
NGC5957  &  B  &  F  & $ -0.113 \,( 0.013 )$  & $ -0.091 \,( 0.006 )$  & $ 8.662 \,( 0.012 )$  & $ 8.643 \,( 0.006 )$  & $ 0.050 \,( 0.002 )$  & $ 0.050 \,( 0.001 )$ \\
NGC6004  &  B  &  G  & $ -0.039 \,( 0.006 )$  & $ -0.054 \,( 0.004 )$  & $ 8.604 \,( 0.006 )$  & $ 8.596 \,( 0.005 )$  & $ 0.054 \,( 0.002 )$  & $ 0.057 \,( 0.001 )$ \\
NGC6063  &  U  &  F  & $ -0.104 \,( 0.008 )$  & $ -0.144 \,( 0.004 )$  & $ 8.605 \,( 0.009 )$  & $ 8.628 \,( 0.005 )$  & $ 0.053 \,( 0.002 )$  & $ 0.056 \,( 0.001 )$ \\
NGC6941  &  B  &  G  & $ -0.035 \,( 0.007 )$  & $ -0.059 \,( 0.005 )$  & $ 8.578 \,( 0.006 )$  & $ 8.598 \,( 0.006 )$  & $ 0.072 \,( 0.002 )$  & $ 0.069 \,( 0.001 )$ \\
NGC7466  &  U  &  F  & $ -0.090 \,( 0.009 )$  & $ -0.068 \,( 0.006 )$  & $ 8.572 \,( 0.009 )$  & $ 8.554 \,( 0.007 )$  & $ 0.043 \,( 0.002 )$  & $ 0.049 \,( 0.002 )$ \\
NGC7489  &  U  &  F  & $ -0.223 \,( 0.005 )$  & $ -0.171 \,( 0.003 )$  & $ 8.622 \,( 0.007 )$  & $ 8.590 \,( 0.004 )$  & $ 0.049 \,( 0.001 )$  & $ 0.062 \,( 0.001 )$ \\
NGC7591  &  B  &  G  & $ -0.071 \,( 0.006 )$  & $ -0.062 \,( 0.004 )$  & $ 8.614 \,( 0.009 )$  & $ 8.594 \,( 0.004 )$  & $ 0.062 \,( 0.001 )$  & $ 0.064 \,( 0.001 )$ \\
NGC7653  &  U  &  F  & $ -0.115 \,( 0.004 )$  & $ -0.093 \,( 0.004 )$  & $ 8.636 \,( 0.005 )$  & $ 8.608 \,( 0.004 )$  & $ 0.050 \,( 0.001 )$  & $ 0.051 \,( 0.001 )$ \\
NGC7716  &  B  &  F  & $ -0.043 \,( 0.006 )$  & $ -0.038 \,( 0.003 )$  & $ 8.534 \,( 0.006 )$  & $ 8.528 \,( 0.004 )$  & $ 0.041 \,( 0.001 )$  & $ 0.049 \,( 0.001 )$ \\
NGC7819  &  B  &  F  & $ -0.139 \,( 0.006 )$  & $ -0.093 \,( 0.005 )$  & $ 8.621 \,( 0.008 )$  & $ 8.581 \,( 0.005 )$  & $ 0.060 \,( 0.002 )$  & $ 0.071 \,( 0.002 )$ \\
UGC00005  &  U  &  F  & $ -0.054 \,( 0.007 )$  & $ -0.068 \,( 0.004 )$  & $ 8.590 \,( 0.008 )$  & $ 8.599 \,( 0.005 )$  & $ 0.041 \,( 0.002 )$  & $ 0.048 \,( 0.002 )$ \\
UGC02311  &  B  &  G  & $ -0.070 \,( 0.006 )$  & $ -0.051 \,( 0.005 )$  & $ 8.580 \,( 0.009 )$  & $ 8.562 \,( 0.008 )$  & $ 0.045 \,( 0.001 )$  & $ 0.051 \,( 0.001 )$ \\
UGC04195  &  B  &  F  & $ -0.090 \,( 0.007 )$  & $ -0.124 \,( 0.005 )$  & $ 8.628 \,( 0.007 )$  & $ 8.652 \,( 0.006 )$  & $ 0.056 \,( 0.002 )$  & $ 0.056 \,( 0.002 )$ \\
UGC04262  &  U  &  F  & $ -0.166 \,( 0.008 )$  & $ -0.182 \,( 0.006 )$  & $ 8.656 \,( 0.008 )$  & $ 8.659 \,( 0.006 )$  & $ 0.084 \,( 0.002 )$  & $ 0.079 \,( 0.002 )$ \\
UGC07012  &  U  &  F  & $ -0.152 \,( 0.072 )$  & $ -0.112 \,( 0.007 )$  & $ 8.510 \,( 0.046 )$  & $ 8.498 \,( 0.006 )$  & $ 0.022 \,( 0.002 )$  & $ 0.068 \,( 0.001 )$ \\
UGC08781  &  B  &  G  & $ +0.007 \,( 0.010 )$  & $ -0.055 \,( 0.009 )$  & $ 8.495 \,( 0.010 )$  & $ 8.539 \,( 0.009 )$  & $ 0.056 \,( 0.002 )$  & $ 0.060 \,( 0.001 )$ \\
UGC09291  &  B  &  F  & $ -0.199 \,( 0.007 )$  & $ -0.152 \,( 0.005 )$  & $ 8.629 \,( 0.008 )$  & $ 8.602 \,( 0.005 )$  & $ 0.056 \,( 0.002 )$  & $ 0.060 \,( 0.001 )$ \\
UGC09777  &  B  &  F  & $ -0.077 \,( 0.015 )$  & $ -0.020 \,( 0.010 )$  & $ 8.552 \,( 0.013 )$  & $ 8.488 \,( 0.011 )$  & $ 0.071 \,( 0.005 )$  & $ 0.074 \,( 0.003 )$ \\
UGC09842  &  B  &  G  & $ -0.103 \,( 0.012 )$  & $ -0.140 \,( 0.008 )$  & $ 8.590 \,( 0.013 )$  & $ 8.610 \,( 0.008 )$  & $ 0.060 \,( 0.002 )$  & $ 0.062 \,( 0.001 )$ \\
UGC12224  &  U  &  F  & $ -0.173 \,( 0.013 )$  & $ -0.122 \,( 0.006 )$  & $ 8.640 \,( 0.011 )$  & $ 8.588 \,( 0.006 )$  & $ 0.054 \,( 0.002 )$  & $ 0.056 \,( 0.001 )$ \\
UGC12816  &  U  &  F  & $ -0.094 \,( 0.008 )$  & $ -0.070 \,( 0.004 )$  & $ 8.506 \,( 0.008 )$  & $ 8.485 \,( 0.005 )$  & $ 0.055 \,( 0.003 )$  & $ 0.068 \,( 0.002 )$ \\
\label{table}
\end{longtable}
\end{longtab}
\newpage

\clearpage
\twocolumn

\end{document}